\documentstyle[12pt]{article}
\renewcommand{\theequation}{\arabic{section}.\arabic{equation}}
\def\case#1#2{{\textstyle{#1\over #2}}}

\title{\hfill{\normalsize ULB/229/CQ/97/2}\\
\vspace{1cm}
Generalized Morse potential: symmetry and satellite potentials}
\author{A. Del Sol Mesa $^{a,}$\thanks{E-mail:
antonio@sysul1.ifisicacu.unam.mx}\ ,
C. Quesne $^{b,}$\thanks{Directeur de recherches FNRS; E-mail:
cquesne@ulb.ac.be}\ ,
Yu. F. Smirnov $^{c,}$\thanks{E-mail: smirnov@xochitl.nuclecu.unam.mx}\\
{\small $^a$ Instituto de F\'{\i}sica, UNAM, Apdo Postal 20-364, 01000
M\'exico DF, Mexico} \\
{\small $^b$ Physique Nucl\'eaire Th\'eorique et Physique Math\'ematique,
Universit\'e Libre de Bruxelles,} \\
{\small Campus de la Plaine CP229, Boulevard
du Triomphe, B-1050 Brussels, Belgium} \\
{\small $^c$ Instituto de Ciencias Nucleares, UNAM, Apdo Postal 70-543, 04510
M\'exico DF, Mexico}}
\date{ }
\begin{document}
\baselineskip=22pt plus 1pt minus 1pt
\maketitle

\begin{abstract}
We study in detail the bound state spectrum of the generalized Morse
potential~(GMP), which was proposed by Deng and Fan as a potential function for
diatomic molecules. By connecting the corresponding Schr\"odinger equation with
the Laplace equation on the hyperboloid and the Schr\"odinger equation for the
P\"oschl-Teller potential, we explain the exact solvability of the problem by an
$so(2,2)$ symmetry algebra, and obtain an explicit realization of the latter as
$su(1,1) \oplus su(1,1)$. We prove that some of the $so(2,2)$ generators connect
among themselves wave functions belonging to different GMP's (called satellite
potentials). The conserved quantity is some combination of the potential
parameters instead of the level energy, as for potential algebras. Hence,
$so(2,2)$
belongs to a new class of symmetry algebras. We also stress the usefulness
of our
algebraic results for simplifying the calculation of Frank-Condon factors for
electromagnetic transitions between rovibrational levels based on different
electronic states.
\end{abstract}

\bigskip
\hspace*{0.3cm}
Short title: Generalized Morse potential

\bigskip
\hspace*{0.3cm}
PACS: 03.65.Fd, 33.10.Gx, 33.10.Cs

\bigskip
\hspace*{0.3cm}
To be published in J. Phys. A

\newpage
%
%
\section{Introduction}
\setcounter{equation}{0}
It is well known that the factorization method of Infeld and Hull~\cite{infeld},
the algebraic approach of Alhassid {\sl et
al\/}~\cite{alhassid83,alhassid86,frank,barut}, and/or the SUSYQM superalgebraic
scheme for shape-invariant potentials~\cite{gendenshtein} allow one to connect
with one another wave functions~$\Psi^{(m)}$ corresponding to a set of
potentials
$V^{(m)}$, $m = 0$, 1, 2,~$\ldots$. The wave functions $\Psi^{(m)}$ satisfy the
Schr\"odinger equations
\begin{equation}
  \left[- \frac{\hbar^2}{2\mu} \frac{d^2}{dx^2} + V^{(m)}(x)\right]
\Psi^{(m)}(x)
  = E \Psi^{(m)}(x)  \label{eq:schrodinger1}
\end{equation}
with the same energy eigenvalue~$E$. In the factorization method, the
connections
between these functions are given by the relations
\begin{equation}
  H^{(\pm)} \Psi^{(m)}(x) = \Psi^{(m\pm1)}(x)    \label{eq:ladder}
\end{equation}
where $H^{(\pm)}$ are $m$-dependent, first-order differential operators.
Explicit expressions for the ladder operators $H^{(\pm)}$ were found
in~\cite{infeld} for various types of potentials.\par
%
%
In the algebraic approach~\cite{alhassid83,alhassid86,frank, barut}, the
$m$-dependence of the ladder operators is eliminated by introducing some
auxiliary
variables, so that the resulting operators coincide with the generators of some
algebra $G$ ($G = su(2)$, $so(2,1)$, $so(2,2)$, etc). A definite irreducible
representation (irrep)~$D^w$ of~$G$ is realized in the space of wave
functions~$\Psi^{(m)}$. The number of interconnected functions~$\Psi^{(m)}$ (and
corresponding potentials~$V^{(m)}$) can be finite ($m \le n-1$, where $n$ is the
dimension of the irrep~$D^w$ of the compact algebra~$G$) or infinite ($D^w$
belongs to the discrete series of unitary irreps of the noncompact
algebra~$G$).\par
%
%
SUSYQM provides an alternative algebraization of the factorization method,
wherein the ladder operators give rise to supercharge operators
$Q^{\pm}$~\cite{gendenshtein,dabrowska}. Together with the supersymmetric
Hamiltonian, the latter generate an $su(1/1)$ superalgebra.\par
%
%
A common feature of the various
approaches~\cite{infeld,alhassid83,alhassid86,frank,barut,gendenshtein,
dabrowska}
is the fact that taking some initial potential $V^{(i)}$, one can construct
a set of
potentials
$V^{(0)}$, $V^{(1)}$,~$\ldots$, $V^{(i-1)}$, $V^{(i+1)}$,~$\ldots$,
associated with it.
If $V^{(i)}$ has a finite number of levels (bound states) $E^{(i)}_s$, $s =
0$, 1,
$\ldots$,~$l$, then the potential $V^{(m)}$ ($m < i$) has the same levels
$E^{(i)}_s$
except for $i-m$ of them, i.e., $s = 0$, 1, $\ldots$,~$l-i+m$. The corresponding
wave functions of such potentials, which may be called {\em satellite
potentials},
are connected by eq.~(\ref{eq:ladder}).
In~\cite{infeld,alhassid83,alhassid86,frank,barut,gendenshtein,dabrowska},
it was
assumed that the  level energy~$E$ is constant
(in~\cite{alhassid83,alhassid86,frank,barut}, $E$ is unambigously connected with
the irrep signature~$w$). Applying the algebra~$G$ generators~$H^{(\pm)}$ on a
state belonging to a potential of this set gives rise to a state belonging
to another
satellite potential. Therefore the algebra~$G$ was called  the potential
algebra~\cite{alhassid83}, or the algebra describing the fixed energy
states of a
family of systems with quantized coupling constants~\cite{barut}.\par
%
%
In the present paper, we consider a potential suggested by Deng and Fan in
1957~\cite{deng} to describe diatomic molecular energy spectra and
electromagnetic transitions, and referred to as the generalized Morse
potential~(GMP). As the standard one~\cite{morse},
\begin{equation}
  V(x)=D{[1-e^{-a(x-x_{e})}]}^2
\end{equation}
where $D$, $a$, and $x_{e}$ are some constants, it has the advantage of
automatically including some anharmonicity effects, while admitting an exact
analytical solution, allowing one to easily calculate various molecular
characteristics. In contrast, it has none of the three well-known
limitations of the
Morse potential, namely (i) to be defined in the interval $-\infty < x <
\infty$,
including the non-physical region $x < 0$, (ii) to be finite at $x=0$
unlike the true
internuclear interaction, and (iii) to contain only two terms in the Dunham
expansion for the energies~$E_n$ with respect to $(n+1/2)$,
\begin{equation}
  E_{n}=\sqrt{\frac{2D}{\mu}} a \hbar (n+1/2)- \frac{a^{2} \hbar^{2}}{2 \mu}
  (n+1/2)^2   \label{eq:Morse-eigenvalue}
\end{equation}
whereas higher order terms are needed for a good description  of the
vibrational levels of some diatomic molecules~\cite{deng}. The GMP therefore
provides an interesting alternative to some recent attempts to eliminate the
Morse potential defects, based on computer calculations~\cite{nicholls}.\par
%
%
Our aim is to explain the exact solvability of the~GMP by determining its
symmetry
algebra. For such a purpose, it will prove convenient to relate the
Schr\"odinger
equation for the~GMP to other equations with known symmetry algebras, namely
the Laplace equation on the hyperboloid~\cite{smirnov}, and the
Schr\"odinger equation for the P\"oschl-Teller potential~(PTP)~\cite{barut}.\par
%
%
The GMP symmetry algebra, resulting from such connections, will turn out to be
distinct from the potential algebras, and to belong to a new class of symmetry
algebras, wherein the operators (generators)~$H^{(\pm)}$ connect with one
another
wave functions~$\Psi^{(m)}(x)$ satisfying the equations
\begin{equation}
  \left[- \frac{\hbar^2}{2\mu} \frac{d^2}{dx^2} + V^{(m)}(x)\right]
\Psi^{(m)}(x)
  = E^{(m)} \Psi^{(m)}(x)
\end{equation}
with $m$-dependent eigenvalues $E^{(m)}$. Contrary to the case of
eq.~(\ref{eq:schrodinger1}), the conserved (constant) quantity for the set of
satellite potentials $V^{(0)}$, $V^{(1)}$, $\ldots$, $V^{(m)}$,~$\ldots$
will not be
the level energy $E$, but some combination $f(a_m)$ of the potential parameters
$a_m$:
\begin{equation}
  f(a_m) = \mbox{constant}.   \label{eq:f}
\end{equation}
\par
%
%
One may still consider some first-order differential operators $H^{(\pm)}$
connecting the wave functions of neighbouring satellite potentials, and
satisfying eq.~(\ref{eq:ladder}). They are similar to those of the factorization
method~\cite{infeld}, except for the fact that they do not conserve the energy
eigenvalue $E$. Under certain conditions, which are fulfilled for the~GMP,
they may
generate some algebra~$G$, associated with the satellite potentials. In this
algebraic case, the wave functions $\Psi^{(m)}(x)$, $m = 0$, 1,~$\ldots$, form a
basis of an irrep~$D^w$ of the algebra~$G$ and its signature~$w$ is
connected with
the constant $f(a_m)$ of eq.~(\ref{eq:f}). Such an approach therefore
allows one to
construct a set of satellite potentials $V^{(m)}$ ($m = 0$, 1,
2,~$\ldots$), which are
different in the general case from potentials produced by the factorization
method~\cite{infeld}, by potential
algebras~\cite{alhassid83,alhassid86,frank,barut}, or by
SUSYQM~\cite{gendenshtein,dabrowska}.\par
%
%
This paper is organized as follows. In section~2, the GMP is reviewed, and
its bound
state spectrum and corresponding normalized wave functions are given. Its
connections with the Laplace operator on the hyperboloid and with the
P\"oschl-Teller potential~(PTP) are established in section~3, and used in
section~4 to construct a symmetry algebra $so(2,2)$ related to the new type of
satellite potentials. Some concluding remarks are given in section~5.\par
%
%
\section{The generalized Morse potential}
\setcounter{equation}{0}
The GMP, introduced by Deng and Fan~\cite{deng}, and related to the Manning-Rosen
potential~\cite{infeld,manning} (also called Eckart potential by some
authors~\cite{dabrowska}), is defined by
\begin{equation}
  V(r)=D \left[1-\frac{b}{e^{a r}-1}\right]^2 \qquad b=e^{a r_{e}}-1
\label{eq:GMP}
\end{equation}
where $0 \le r<\infty$, and $D$, $b$, $a$ are some parameters regulating
the depth,
the position of the minimum~$r_e$, and the radius of the potential. Notice
that it
is defined on the same range, and has the same behaviour for~$r\to 0$, as
the true
internuclear potential in diatomic molecules.\par
%
%
In the dimensionless variable~$x = ar$ ($0\le x < \infty$), the Schr\"odinger
equation for the GMP can be written as
\begin{equation}
  \left( - \frac{d^2}{dx^2} + v(x)\right) \psi(x) = \epsilon \psi(x) \qquad
  v(x) \equiv  k \left(1 - \frac{b}{e^x-1}\right)^2 \label{eq:schrodinger-GMP}
\end{equation}
where $\Psi(r) = \sqrt{a} \psi(x)$, and
\begin{equation}
  k\equiv \frac{2\mu D}{a^2 \hbar^2} \qquad \epsilon \equiv \frac{2\mu E}{a^2
  \hbar^2}.
\end{equation}
As was shown in~\cite{deng}, it is solvable in an analytical way. The
introduction
of a new variable~$y$ and a new function~$F(y)$, defined by
\begin{eqnarray}
  y & = &\left(e^x-1\right)^{-1} \label{eq:y} \\
  \psi(x) & = & \Phi(y) = \frac{y^{\alpha}}{(1+y)^{\beta}} F(y)
\label{eq:Phi}
\end{eqnarray}
respectively, indeed transforms eq.~(\ref{eq:schrodinger-GMP}) into the
following equation
\begin{equation}
  y (1+y) \frac{d^{2} F}{d y^{2}} + [2 (\alpha-\beta+1) y +(2 \alpha +1)]
  \frac{d F}{d y} + [(\alpha-\beta)^{2} + (\alpha-\beta) + C] F(y)=0
  \label{eq:hypergeometric-eq}
\end{equation}
equivalent to the hypergeometric one, provided $\alpha$, $\beta$, and $C$ are
chosen so as to satisfy the relations
\begin{eqnarray}
  \alpha & = & \sqrt{k-\epsilon} = \sqrt{\frac{2 \mu (D-E)}{a^{2} \hbar^{2}}}
           \label{eq:alpha} \\
  \beta & = & \sqrt{\alpha^2 + k b (b+2)} = \sqrt{k(b+1)^2-\epsilon} =
\sqrt{\frac{2
           \mu}{a^{2} \hbar^{2}} [D(b+1)^{2} - E]} \label{eq:beta} \\
  C & = & - k b^2 = -\frac{2\mu}{a^{2} \hbar^{2}} b^{2} D. \label{eq:C}
\end{eqnarray}
A solution of~(\ref{eq:hypergeometric-eq}) is $F(y) = {}_2F_1(d, e;
2\alpha+1; -y)$,
where
\begin{equation}
  d = \alpha-\beta+l \qquad e = \alpha-\beta+1-l \qquad l \equiv
\case{1}{2} \left(1
  + \sqrt{1-4C}\right) = \case{1}{2} \left(1 + \sqrt{1+4kb^2}\right).
  \label{eq:d-e-l}
\end{equation}
\par
%
%
Bound states correspond to those functions $F(y)$ that reduce to
polynomials, i.e.,
for which $d=-n$, where $n$ is some nonnegative integer. From
eqs.~(\ref{eq:beta})
and~(\ref{eq:d-e-l}), it is then clear that $\alpha$ and $\beta$ are
$n$-dependent, and solutions of the equations
\begin{eqnarray}
  \beta_n - \alpha_n & = & n + l \label{eq:eq-alpha-beta-1} \\
  \beta_n^2 - \alpha_n^2 & = & kb(b+2). \label{eq:eq-alpha-beta-2}
\end{eqnarray}
One finds
\begin{equation}
  \alpha_n = \frac{1}{2} \left(\frac{kb(b+2)}{n+l} - n - l\right) \qquad
  \beta_n = \frac{1}{2} \left(\frac{kb(b+2)}{n+l} + n + l\right).
  \label{eq:alpha-beta}
\end{equation}
From eq.~(\ref{eq:alpha}), it follows that $\epsilon_n$ can be expressed in
terms
of~$\alpha_n$ as
\begin{equation}
  \epsilon_n = k - \alpha_n^2.   \label{eq:epsilon}
\end{equation}
\par
%
%
So, we conclude that the energy eigenvalues are given by
\begin{equation}
  E_{n}=D - \frac{a^2\hbar^2}{8 \mu} \left(n + l -\frac{b(b+2)k}{n+l}\right)^2
  \label{eq:eigenvalue}
\end{equation}
and that the corresponding eigenfunctions are
\begin{equation}
  \Psi_{n}(r) = N_n y^{\alpha_n} (1+y)^{-\beta_n} {}_2F_1(-n, -n+1-2l;
2\alpha_n+1;
  -y) \qquad y = \left(e^{ar}-1\right)^{-1}   \label{eq:eigenfunction}
\end{equation}
where $N_n$ is some normalization coefficient, determined by the condition
\begin{equation}
  \int_0^{\infty} dr\, |\Psi_n(r)|^2 = \int_0^{\infty} dx\, |\psi_n(x)|^2 =
  \int_0^{\infty} dy\, [y(1+y)]^{-1} |\Phi_n(y)|^2 = 1.
\end{equation}
In eqs.~(\ref{eq:eigenvalue})
and~(\ref{eq:eigenfunction}), the quantum number~$n$ takes a finite set of
values
\begin{equation}
  n=0,1,2,\ldots,n_{max} \qquad n_{max} \equiv \sqrt{k b (b+2)} - l
\end{equation}
where the quantity defining $n_{max}$ is assumed integer (otherwise one has
to take its integer part).\par
%
%
It is not a trivial matter to derive a closed expression for the normalization
coefficient~$N_n$ of the eigenfunctions~(\ref{eq:eigenfunction}), which was not
given in~\cite{deng}. As shown in the appendix, this can most easily be done by
using SUSYQM techniques. The result can be rewritten as
\begin{equation}
  N_n = \left(\frac{a (\alpha_n+n+l) \Gamma(2\alpha_n+n+1)
  \Gamma(2\alpha_n+n+2l)}{n!\, (n+l) \Gamma(2\alpha_n) \Gamma(2\alpha_n+1)
  \Gamma(n+2l)}\right)^{1/2}.   \label{eq:normalization}
\end{equation}
\par
%
%
Making an expansion of~(\ref{eq:eigenvalue}) in terms of powers of
$(n+1/2)$, we
get~\cite{deng}
\begin{equation}
  E_n = \epsilon(0) + \epsilon(1) (n+1/2) - \epsilon(2) (n+1/2)^{2} +
\epsilon(3)
  (n+1/2)^{3} - \cdots   \label{eq:approximation}
\end{equation}
where $\epsilon(0)$, $\epsilon(1)$, $\epsilon(2)$, and~$\epsilon(3)$ are
coefficients depending on the parameters of the potential function. This means
that the GMP includes terms of arbitrary order in the Dunham expansion. It
is easy to
verify that all corrections $\epsilon(k) \rightarrow 0$ $(k \geq 3)$ whenever $b
\rightarrow \infty$. Therefore, in this limit, the
eigenvalues~(\ref{eq:eigenvalue})
coincide with those of the Morse potential, given in
eq.~(\ref{eq:Morse-eigenvalue}).\par
%
%
On the other hand, using the well-known limit relation between Gauss and
confluent hypergeometric functions ${}_2F_1(a,b,c;z/b) \rightarrow
{}_1F_1(a,c;z)$, when $b \rightarrow \infty$ (or $r_{e} \gg
1$)~\cite{slater}, and considering the region $r \gg 1$, we obtain that the GMP
eigenfunctions~(\ref{eq:eigenfunction}) can be reduced to those of the Morse
potential.\par
%
%
The interrelation  between the Morse potential and the GMP is illustrated
in Fig.1.
It is clear that for rather large values of $r_{e}$ ($a=1$), these
potentials are very
close to each other in the regions $r \sim r_{e}$ and $r > r_{e}$. However,
they are
very  different at $r \sim 0$. If both potentials are rather deep ($D \gg
1$), they
could be well approximated, in the region $r \sim r_{e}$, by a harmonic
oscillator
potential with frequency $\hbar \omega=\epsilon(1)$ (see
eq.~(\ref{eq:approximation})).\par
%
%
Usually, the existence of an exact analytical solution of the Schr\"odinger
equation for some system can be explained by the fact that the corresponding
Hamiltonian has some symmetry algebra. In the next two sections, we shall
proceed
to determine the latter for the GMP.\par
%
%
\section{Connections with the Laplace operator on the hyperboloid and the
P\"oschl-Teller potential}
\setcounter{equation}{0}
As well known, in order to find the symmetry algebra related to a
one-dimensional Schr\"odinger equation, it is  useful to map the latter into a
problem in a higher-dimensional space. For instance, by using some
transformation,
Alhassid {\em et al}~\cite{alhassid83} were able to map the Schr\"odinger
equation
for the Morse potential into a two-dimensional harmonic oscillator.\par
%
%
Following this line of thought, we will embed our one-dimensional GMP~problem
into a three-dimensional space. Namely, we will show below that the  Schr\"odinger
equation for the GMP can be connected with the Laplace-Beltrami-Casimir one on
the four-dimensional hyperboloid defined by
\begin{equation}
  x_{1}^{2}+x_{2}^{2}-x_{3}^{2}-x_{4}^{2}=\rho^{2}>0
\end{equation}
whose symmetry algebra is $so(2,2)$.\par
%
Using the Casimir operator of this algebra, we will derive the discrete
spectrum of
the GMP, and show that the corresponding wave functions can be connected
with the
eigenfunctions of the Laplace operator on the  hyperboloid, or,
equivalently, with
those of the Schr\"odinger equation for the P\"oschl-Teller potential~(PTP),
\begin{equation}
  V_{P T}=\frac{\lambda^2}{\sinh^2 \theta} -\frac{{\overline \lambda}^2}
{\cosh^2
  \theta} \qquad 0 \leq \theta < \infty.
\end{equation}
We will therefore conclude that the exact solvability of the Schr\"odinger
equation for the GMP is explained by its connection with the $so(2,2)$
algebra.\par
%
%
Let us analyze the symmetry problem for the GMP in detail. We start by
considering
the four-dimensional Minkowski space determined by the
relations~\cite{smirnov,raczka}
\begin{eqnarray}
  x_{1} & = & \rho\, \cosh\theta \cos\varphi \qquad x_{2}=\rho\, \cosh\theta
        \sin\varphi \nonumber \\
  x_{3} & = & \rho\, \sinh\theta \cos\phi \qquad x_{4}=\rho\, \sinh\theta
        \sin\phi   \label{eq:minkowski}
\end{eqnarray}
where $0 \leq \theta < \infty$, $0 \leq \varphi < 2 \pi$, $0 \leq \phi <
2\pi$, $\rho
\geq 0$. In such coordinates, the Laplace-Beltrami-Casimir operator takes the
form
\begin{equation}
  \Delta_{\Omega} = - \frac{1}{{\sinh \theta} {\cosh \theta}} \frac{\partial}
  {\partial \theta} \left({\cosh \theta} {\sinh \theta} \frac{\partial}
{\partial
  \theta}\right) + \frac{1}{\cosh^2 \theta} \frac{\partial^2}{\partial\varphi^2}
  - \frac{1}{\sinh^2 \theta} \frac{\partial^2}{\partial\phi^2}.
\end{equation}
Its eigenvalues can be written as
\begin{equation}
  \lambda=-L (L+2) \qquad L=0,1,2,\ldots,L_{max}
\end{equation}
and its eigenfunctions $\Psi_{L m_{1} m_{2}}(\theta, \varphi, \phi)$ can be
factorized as follows
\begin{equation}
  \Psi_{L m_{1} m_{2}}(\theta, \varphi, \phi) = e^{i m_{1} \varphi}
\psi_{L m_{1}
  m_{2}}(\theta) e^{i m_{2} \phi}
\end{equation}
where $\psi_{L m_{1} m_{2}}(\theta)$ satisfies the equation
\begin{eqnarray}
  \left[-\frac{\partial^2}{\partial \theta^2}-(\tanh \theta+\coth \theta)
         \frac{\partial}{\partial \theta}-\frac{m_{1}^2}{\cosh^2 \theta}+
         \frac{m_{2}^2}{\sinh^2 \theta}\right] \psi_{L m_{1} m_{2}}(\theta)
         \nonumber \\[0.2cm]
 \mbox{} = -L (L+2) \psi_{L m_{1} m_{2}}(\theta).
         \label{eq:Laplace-eq}
\end{eqnarray}
\par
%
%
We notice that according to the choice of coordinates~(\ref{eq:minkowski}), the
problem of determining the eigenvalues and eigenfunctions of the
Laplace-Beltrami-Casimir operator corresponds to the reduction
\begin{equation}
  so(2,2) \supset so_{1}(2) \oplus so_{2}(2)
\end{equation}
of the $so(2,2)$ algebra into its subalgebras $so_1(2)$ and $so_2(2)$,
generated by the operators $-i \partial/\partial\varphi$ and
$-i \partial/\partial\phi$, respectively. The eigenvalues
\begin{equation}
  m_1, m_2 = 0, \pm 1,\pm 2, \ldots
\end{equation}
of the latter enumerate the irreps of these subalgebras. The quantum number
\begin{equation}
  L = 0, 1, 2, \ldots
\end{equation}
characterizes the irrep of the algebra $so(2,2)$, belonging to the discrete
series of its most degenerate unitary irreps~\cite{raczka}.\par
%
%
It is well known from the general theory of $so(p,q)$ irreps~\cite{raczka},
that in a
given irrep $D^L$ of $so(2,2)$, the admissible values of the quantum
numbers $m_1$
and $m_2$ are given by the condition
\begin{equation}
  |m_1| - |m_2| = L + 2 + 2n \qquad n = 0, 1, 2, \ldots.
\end{equation}
Therefore, if $m_1$ and $m_2$ are fixed in eq.~(\ref{eq:Laplace-eq}), then
\begin{equation}
  L = L_{max}, L_{max}-2, \ldots, 1 \mbox{\ or\ } 0
\end{equation}
depending on
\begin{equation}
  L_{max} = |m_1|- |m_2| -2
\end{equation}
being odd or even, respectively. Thus, the total number of possible
$L$~values is
\begin{equation}
  \nu_{max} = \case{1}{2}\left(|m_1|-|m_2|\right) \quad \mbox{or} \quad
  \case{1}{2}\left(|m_1|-|m_2|-1 \right)   \label{eq:numax}
\end{equation}
for $L_{max}$ even or odd, respectively.\par
%
%
Before bringing the discussion further, it is interesting to make the following
substitution
\begin{equation}
  \psi_{L_{m_1m_2}}(\theta) = \cosh^{-1/2}\theta \sinh^{-1/2}\theta\,
\chi(\theta)
\end{equation}
transforming the Laplace equation~(\ref{eq:Laplace-eq}) into the Schr\"odinger
equation for the PTP
\begin{equation}
  \left[ -\frac{d^2}{d\theta^2} + \frac{(m_2^2 - 1/4)}{\sinh^2\theta} -
  \frac{(m_1^2 - 1/4)}{\cosh^2\theta}\right] \chi(\theta) =
\overline{\epsilon}\,
  \chi(\theta)   \label{eq:schrodinger-PTP}
\end{equation}
where
\begin{equation}
  \overline{\epsilon} = -(L+1)^2 = -(|m_1| - |m_2| -1 - 2n)^2.
  \label{eq:PTP-eigenvalue}
\end{equation}
\par
%
%
Equation~(\ref{eq:numax}) shows the number of bound states in the PTP for fixed
amplitudes $m_1$ and $m_2$ of its attractive and  repulsive parts. If
\begin{equation}
  |m_2| - |m_1| \geq -1
\end{equation}
then there are no bound states.\par
%
%
Let us now establish the connection of the GMP problem with the $so(2,2)$
Laplace
equation and the Schr\"odinger equation~(\ref{eq:schrodinger-PTP}). It is easy
to check that the change of variable and of function
(\ref{eq:y}),~(\ref{eq:Phi}),
followed by the transformation
\begin{equation}
  y = \sinh^2\theta \qquad \Phi(y) = y^{-1/4}(1+y)^{-1/4} \Xi(y) \qquad \Xi(y) =
  \chi(\theta)
  \label{eq:Laplace-GMP}
\end{equation}
maps the Schr\"odinger equation for the GMP into that corresponding to the PTP,
given in eq.~(\ref{eq:schrodinger-PTP}). Therefore, we can use the results
obtained
for the latter to find the level spectrum and corresponding eigenfunctions
of the
GMP. For such purpose, we should identify
\begin{equation}
  |m_1| = 2 \beta \qquad |m_2| = 2 \alpha \qquad \overline{\epsilon} = 4 C-1 =
  - (2l-1)^2.   \label{eq:PTP-GMP}
\end{equation}
It then follows from eq.~(\ref{eq:PTP-eigenvalue}) that
\begin{equation}
  (2l-1)^2 = [ 2(\beta -\alpha) - 1- 2n]^2.  \label{eq:PTP-eigenvaluebis}
\end{equation}
Hence, we directly obtain eq.~(\ref{eq:eq-alpha-beta-1})\footnote{The other
solution of eq.~(\ref{eq:PTP-eigenvaluebis}) can easily be shown to violate the
condition $\alpha>0$.}, and from it the GMP spectrum~(\ref{eq:eigenvalue}).\par
%
%
Thus, we have shown that the GMP eigenvalue spectrum can be obtained by
using the
Laplace-Beltrami-Casimir operator for the $so(2,2)$ algebra. The
eigenfunctions of
the Laplace operator (or, equivalently, of the PTP Hamiltonian) are related
to those
of the GMP through eq.~(\ref{eq:Laplace-GMP}). We have therefore
established that
$so(2,2)$ is the algebra responsible for the GMP exact solvability. In the next
section, we shall proceed to construct its generators, and to study their
action on
the GMP wave functions.\par
%
%
\section{Symmetry algebra associated with the generalized Morse potential}
\setcounter{equation}{0}
According to the analysis of the previous section, by interchanging the roles of
energy and potential parameters, our original Schr\"odinger equation for the GMP
can be mapped into the Schr\"odinger equation for the PTP
(\ref{eq:schrodinger-PTP}), provided we make the
identifications~(\ref{eq:PTP-GMP}).\par
%
%
In terms of the variable~$y$ defined in~(\ref{eq:Laplace-GMP}), the
(unnormalized)
solutions of the PTP~Schr\"odinger equation can be written as
\begin{equation}
  \Xi(y) = y^{\alpha +1/4} (1+y)^{1/4-\beta } {}_2F_1(-n,-n+1-2l;2\alpha+1;-y)
  \label{eq:eigenfunction-PTP}
\end{equation}
where use has been made of eqs.~(\ref{eq:Phi}), (\ref{eq:eigenfunction}),
and~(\ref{eq:Laplace-GMP}). Let us define
\begin{equation}
  | m_1| = m-g = 2\beta \qquad |m_2| = m+g = 2\alpha \qquad \mbox{or} \qquad m =
  \alpha+\beta \qquad g = \alpha-\beta   \label{eq:m-g}
\end{equation}
and make this substitution in eq.~(\ref{eq:eigenfunction-PTP}). Taking
eq.~(\ref{eq:eq-alpha-beta-1}) into account, we obtain
\begin{equation}
  \Xi^{(l)}_{m,g}(y) = y^{(g+m+1/2)/2} (1+y)^{(g-m+1/2)/2}
{}_2F_1(g+l,g+1-l;m+g+1;
  -y)
\end{equation}
where for reasons that will soon become clear, we introduced upper and lower
indices~$(l)$, and $m$, $g$, respectively.\par
%
%
We are now able to show that the algebra $so(2,2)$ associated with the GMP
can be
explicitly represented as $su(1,1) \oplus su(1,1)$. In fact, according to
Barut {\it et
al}~\cite{barut}, for the PTP we can write down two sets of generators
$\overline{G}^+$, $\overline{G}^-$,~$\overline{G}_3$, and $\overline{M}^+$,
$\overline{M}^-$,~$\overline{M}_3$, each corresponding to an $su(1,1)$ algebra,
i.e., satisfying the relations
\begin{eqnarray}
  \left[\overline{G}_3, \overline{G}_{\pm}\right] & = & \pm \overline{G}_{\pm}
         \qquad \left[\overline{G}_+, \overline{G}_-\right] = - 2\overline{G}_3
         \nonumber \\
  \left[\overline{M}_3, \overline{M}_{\pm}\right] & = & \pm \overline{M}_{\pm}
         \qquad \left[\overline{M}_+, \overline{M}_-\right] = - 2\overline{M}_3.
\end{eqnarray}
In terms of the variable~$y$, and of two auxiliary variables~$\xi$, $\eta
\in [0,
2\pi)$, they are given by
\begin{eqnarray}
  \overline{G}^{\pm} & = & -\frac{1}{2} e^{\pm i\eta} \Biggl[\mp 2 \sqrt{y(1+y)}
         \frac{\partial}{\partial y} + \sqrt{\frac{1+y}{y}} \left(-i
\left(\frac{\partial}
         {\partial\xi} + \frac{\partial}{\partial\eta}\right) \pm
\frac{1}{2}\right)
         \nonumber \\
  & & \mbox{} + \sqrt{\frac{y}{1+y}} \left(i
\left(\frac{\partial}{\partial\xi} -
         \frac{\partial}{\partial\eta}\right) \pm \frac{1}{2}\right)\Biggr]
\nonumber
         \\
  \overline{G}_3 & = & -i\frac{\partial}{\partial\eta}
\end{eqnarray}
and
\begin{eqnarray}
  \overline{M}^{\pm} & = & \frac{1}{2} e^{\pm i\xi} \Biggl[\mp 2 \sqrt{y(1+y)}
         \frac{\partial}{\partial y} + \sqrt{\frac{1+y}{y}} \left(-i
\left(\frac{\partial}
         {\partial\xi} + \frac{\partial}{\partial\eta}\right) \pm
\frac{1}{2}\right)
         \nonumber \\
  & & \mbox{} + \sqrt{\frac{y}{1+y}} \left(-i
\left(\frac{\partial}{\partial\xi} -
         \frac{\partial}{\partial\eta}\right) \pm \frac{1}{2}\right)\Biggr]
\nonumber
         \\
  \overline{M}_3 & = & -i\frac{\partial}{\partial\xi}
\end{eqnarray}
respectively.\par
%
%
Applying them on the extended PTP~wave functions
\begin{equation}
  \overline{\Xi}^{(l)}_{m,g}(\xi,y,\eta) = e^{i m\xi} \Xi^{(l)}_{m,g}(y)
e^{i g\eta},
\end{equation}
we obtain after some calculations using well-known properties of the
hypergeometric function~\cite{slater}
\begin{eqnarray}
  \overline{G}^+\, \overline{\Xi}^{(l)}_{m,g}(\xi,y,\eta) & = &
          \frac{(l+g)(l-g-1)}{m+g+1}\, \overline{\Xi}^{(l)}_{m,g+1}(\xi,y,\eta)
          \nonumber \\
  \overline{G}^-\, \overline{\Xi}^{(l)}_{m,g}(\xi,y,\eta) & = & - (m+g)\,
          \overline{\Xi}^{(l)}_{m,g-1}(\xi,y,\eta) \nonumber \\
  \overline{G}_3\, \overline{\Xi}^{(l)}_{m,g}(\xi,y,\eta) & = & g\,
          \overline{\Xi}^{(l)}_{m,g}(\xi,y,\eta)   \label{eq:Gbar-action}
\end{eqnarray}
and
\begin{eqnarray}
  \overline{M}^+\, \overline{\Xi}^{(l)}_{m,g}(\xi,y,\eta) & = &
          \frac{(m+l)(m-l+1)}{m+g+1}\, \overline{\Xi}^{(l)}_{m+1,g}(\xi,y,\eta)
          \nonumber \\
  \overline{M}^-\, \overline{\Xi}^{(l)}_{m,g}(\xi,y,\eta) & = & (m+g)\,
          \overline{\Xi}^{(l)}_{m-1,g}(\xi,y,\eta) \nonumber \\
  \overline{M}_3\, \overline{\Xi}^{(l)}_{m,g}(\xi,y,\eta) & = & m\,
          \overline{\Xi}^{(l)}_{m,g}(\xi,y,\eta).   \label{eq:Mbar-action}
\end{eqnarray}
\par
%
%
From eq.~(\ref{eq:Laplace-GMP}), the corresponding operators $G^+$, $G^-$,
$G_3$,
$M^+$, $M^-$,~$M_3$ for the GMP can be written as $G^+ = [y(1+y)]^{-1/4}\,
\overline{G}^+ [y(1+y)]^{1/4}$, and similarly for the other generators. Their
detailed expressions are given by
\begin{eqnarray}
  G^{\pm} & = & -\frac{1}{2} e^{\pm i\eta} \Biggl[\mp 2 \sqrt{y(1+y)}
         \frac{\partial}{\partial y} - i \sqrt{\frac{1+y}{y}}
\left(\frac{\partial}
         {\partial\xi} + \frac{\partial}{\partial\eta}\right)
         + i \sqrt{\frac{y}{1+y}} \left(\frac{\partial}{\partial\xi} -
         \frac{\partial}{\partial\eta}\right)\Biggr] \nonumber \\
  G_3 & = & -i\frac{\partial}{\partial\eta} \nonumber \\
  M^{\pm} & = & \frac{1}{2} e^{\pm i\xi} \Biggl[\mp 2 \sqrt{y(1+y)}
         \frac{\partial}{\partial y} - i \sqrt{\frac{1+y}{y}}
\left(\frac{\partial}
         {\partial\xi} + \frac{\partial}{\partial\eta}\right)
         - i \sqrt{\frac{y}{1+y}} \left(\frac{\partial}{\partial\xi} -
         \frac{\partial}{\partial\eta}\right)\Biggr] \nonumber \\
  M_3 & = & -i\frac{\partial}{\partial\xi}.
\end{eqnarray}
\par
%
%
Defining now extended, normalized GMP~wave functions by
\begin{equation}
  \overline{\Phi}^{(l)}_{m,g}(\xi,y,\eta) = (2\pi)^{-1} e^{i m\xi}
\Phi^{(l)}_{m,g}(y)
  e^{i g\eta},  \label{eq:extended}
\end{equation}
where $\Phi^{(l)}_{m,g}(y)$ is to be identified with the function~$\Phi_n(y)$
obtained from eq.~(\ref{eq:eigenfunction}), we get
\begin{eqnarray}
  G^+ \overline{\Phi}^{(l)}_{m,g}(\xi,y,\eta) & = & - \left(
          \frac{(g+1)(m-g)(m+g)(g+l)(g-l+1)}{g(m-g-1)(m+g+1)}\right)^{1/2}
          \overline{\Phi}^{(l)}_{m,g+1}(\xi,y,\eta) \nonumber \\
  G^- \overline{\Phi}^{(l)}_{m,g}(\xi,y,\eta) & = & - \left(
          \frac{(g-1)(m-g)(m+g)(g-l)(g+l-1)}{g(m-g+1)(m+g-1)}\right)^{1/2}
          \overline{\Phi}^{(l)}_{m,g-1}(\xi,y,\eta) \nonumber \\
  G_3 \overline{\Phi}^{(l)}_{m,g}(\xi,y,\eta) & = & g\,
          \overline{\Phi}^{(l)}_{m,g}(\xi,y,\eta) \label{eq:G-action} \\
  M^+ \overline{\Phi}^{(l)}_{m,g}(\xi,y,\eta) & = & \left(
          \frac{(m-g)(m+g)(m+l)(m-l+1)}{(m-g+1)(m+g+1)}\right)^{1/2}
          \overline{\Phi}^{(l)}_{m+1,g}(\xi,y,\eta) \nonumber \\
  M^- \overline{\Phi}^{(l)}_{m,g}(\xi,y,\eta) & = & \left(
          \frac{(m-g)(m+g)(m-l)(m+l-1)}{(m-g-1)(m+g-1)}\right)^{1/2}
          \overline{\Phi}^{(l)}_{m-1,g}(\xi,y,\eta) \nonumber \\
  M_3 \overline{\Phi}^{(l)}_{m,g}(\xi,y,\eta) & = & m\,
          \overline{\Phi}^{(l)}_{m,g}(\xi,y,\eta).  \label{eq:M-action}
\end{eqnarray}
In deriving eqs.~(\ref{eq:G-action}) and~(\ref{eq:M-action}), we used the
fact that
the action of $G^{\pm}$, $G_3$, $M^{\pm}$,~$M_3$ on the extended, unnormalized
GMP~wave functions is the same as that of $\overline{G}^{\pm}$,
$\overline{G}_3$,
$\overline{M}^{\pm}$,~$\overline{M}_3$ on
$\overline{\Xi}^{(l)}_{m,g}(\xi,y,\eta)$,
as well as the expression~(\ref{eq:normalization}) of the GMP~wave function
normalization coefficient.\par
%
%
We can also write down the expressions for the Casimir operators of each
$su(1,1)$
algebra,
\begin{equation}
  {\cal C}_{su(1,1)_I} = - G^+ G^- + G_3^2 - G_3 \qquad
  {\cal C}_{su(1,1)_{II}} = - M^+ M^- + M_3^2 - M_3.
\end{equation}
From~(\ref{eq:G-action}) and~(\ref{eq:M-action}), it is easy to prove that their
action on the extended wave functions~(\ref{eq:extended}) is
\begin{eqnarray}
  {\cal C}_{su(1,1)_I} \overline{\Phi}^{(l)}_{m,g}(\xi,y,\eta) & = &
          {\cal C}_{su(1,1)_{II}} \overline{\Phi}^{(l)}_{m,g}(\xi,y,\eta) =
          l(l-1) \overline{\Phi}^{(l)}_{m,g}(\xi,y,\eta) \nonumber \\
  & = & - C \overline{\Phi}^{(l)}_{m,g}(\xi,y,\eta).
\end{eqnarray}
The irreducible representations of both $su(1,1)$ algebras are therefore
characterized by~$l$ or, equivalently, by~$C = -kb^2$, or by the
combination $Db^2
/a^2$ of the parameters regulating the depth, the position of the minimum,
and the
radius of the GMP, respectively (see eqs.~(\ref{eq:C})
and~(\ref{eq:d-e-l})).\par
%
%
As we can see from eq.~(\ref{eq:G-action}), the operators~$G^{\pm}$ change $g$
into $g\pm1$, respectively, without changing~$m$. According to the
definitions~(\ref{eq:m-g}) of~$g$ and~$m$, and
eqs.~(\ref{eq:eq-alpha-beta-1}),~(\ref{eq:eq-alpha-beta-2}),
\begin{eqnarray}
  g & = & -l-n \qquad n=0,1,2,\ldots  \label{eq:g} \\
  m & = & \frac{C(b+2)}{gb}  \label{eq:m}
\end{eqnarray}
so that the action of~$G^{\pm}$ forces $n$ and~$b$ to change.\par
%
%
Let us analyse these changes in detail. Since $g\to g\pm 1$, and $C =
-l(l-1)$ is a constant in a given irrep of $su(1,1)_I$, we have that $n\to
n\mp 1$.
On the other hand, $m\to m$ means that
$b$ should change from~$b_g$ to
\begin{equation}
  b_{g\pm1} = \frac{2gb_g}{\left(2g\pm b_g\pm2\right)}.
\end{equation}
From the definition~(\ref{eq:C}) of~$C$, we then obtain $k_g \to
k_{g\pm1}$,  $D_g
\to D_{g\pm1}$,  $a_g \to a_{g\pm1}$, where
\begin{equation}
  k_g b_g^2 = k_{g\pm 1} b_{g\pm 1}^2 \qquad \frac{D_gb_g^2}{a_g^2} =
  \frac{D_{g\pm1}b_{g\pm 1}^2}{a_{g\pm1}^2}.
\end{equation}
\par
%
%
As eq.~(\ref{eq:g}) shows, the $su(1,1)$ irreps associated with~$g$
correspond to
the negative discrete series. The highest-weight vector corresponds to $g=-l$ or
$n=0$, and according to eq.~(\ref{eq:G-action}), it satisfies the relation
\begin{equation}
   G^+ \overline{\Phi}^{(l)}_{m,-l}(\xi,y,\eta)=0
\end{equation}
as it should be. Since the matrix representing~$G^+$ in the basis $\left\{
\overline{\Phi}^{(l)}_{m,g}\right\}$ is not the adjoint of that
representing~$G^-$,
the corresponding irrep is non-unitary.\par
%
%
Let us now analyze the action of the operators $M^{\pm }$, given in
eq.~(\ref{eq:M-action}). In this case $m\to m\pm 1$, and $g$ does not change.
From eqs.~(\ref{eq:g}) and~(\ref{eq:m}), we see that $n$ does not change,
and $b$
changes appropriately from~$b_m$ to
\begin{equation}
  b_{m\pm 1}=\frac{2Cb_m}{2C\pm gb_m}.
\end{equation}
From the constancy of~$C$, we have again $k_m \to k_{m\pm1}$,  $D_m \to
D_{m\pm1}$,  $a_m \to a_{m\pm1}$, where
\begin{equation}
  k_m b_m^2 = k_{m\pm 1} b_{m\pm 1}^2 \qquad \frac{D_mb_m^2}{a_m^2} =
  \frac{D_{m\pm1}b_{m\pm 1}^2}{a_{m\pm1}^2}.
\end{equation}
\par
%
%
As it can be checked from eq.~(\ref{eq:M-action}), the function
$\overline{\Phi}^{(l)}
_{l,g}$, corresponding to~$m=l$, satisfies the relation
\begin{equation}
   M^- \overline{\Phi}^{(l)}_{l,g}(\xi,y,\eta)=0.
\end{equation}
Hence, the $su(1,1)$ irreducible representations associated with~$m$ belong
to the
positive discrete series, i.e.,
\begin{equation}
   m = l+v \qquad v=0,1,2,\ldots.
\end{equation}
They are non-unitary, as already noted for those associated with~$g$.\par
%
%
So as advertised in section~1, we have shown that the GMP $so(2,2)$ symmetry
algebra does not leave invariant the energy eigenvalue of a set of satellite
potentials, but instead the parameter combination~(\ref{eq:f}), where the
function~$f$ is given by $f = D b^2/a^2$, and for the index~$m$ we may use
either
$g$ or~$m$, as defined in eq.~(\ref{eq:m-g}). As a final point, it is worth
stressing
that for the $so(2,2)$ algebra (but not for the corresponding group $SO(2,2)$), the
quantum number~$l$, characterizing its irreducible representations, is not
restricted to integer or half-integer values. Hence our algebraic formalism
may be
used for any real values of the GMP~parameters $D$, $b$, and~$a$, for which
bound
states do exist.\par
%
%
\section{Concluding remarks}
In the present paper, we did study in detail the bound state spectrum of
the GMP,
previously proposed by Deng and Fan~\cite{deng} as a potential function for
diatomic molecules. By connecting the corresponding Schr\"odinger equation with
the Laplace equation on the hyperboloid and the Schr\"odinger equation for the
PTP, we did explain the exact solvability of the problem by an $so(2,2)$
symmetry
algebra, giving rise to a set of satellite potentials of a new type. Such a
symmetry
algebra differs from the well-known potential
algebras~\cite{alhassid83,alhassid86,frank,barut} by the fact that its Casimir
operators are not related to the Hamiltonian as for the latter, but to some
function
of the potential parameters.\par
%
%
It is worth noticing that some algebras with generators simultaneously changing
the energy and the potential parameters, as the GMP $so(2,2)$ symmetry algebra,
did already occur in another context. Some years ago, various attempts have
indeed
been made to combine features of both dynamical and potential algebras by
enlarging the latter with some operators connecting eigenfunctions corresponding
to the same potential parameters, but different energy
eigenvalues~\cite{alhassid83,barut,cq}. The resulting algebras, referred to as
dynamical potential algebras~\cite{cq}, may contain as substructures some
algebras with the above-mentioned characteristics. However these subalgebras
strikingly differ from the GMP symmetry algebra, in the sense that their Casimir
operators are some complicated functions of both the Hamiltonian and the
potential
parameters, instead of the latter only.\par
%
%
The new type of satellite potentials introduced in the present paper may be
physically relevant in the following context. The vibrational potentials $V(r)$
and~$V'(r)$, corresponding to different electronic states $K_e$ and~$K'_e$ of a
diatomic molecule, are in general different. In an analysis of electromagnetic
transitions between rovibrational bands, based on the electronic states
$K_e$ and
$K'_e$ respectively, the corresponding eigenfunctions $\Psi(r)$ and
$\Psi'(r)$ in the
potentials $V(r)$ and $V'(r)$ should be used to calculate the Frank-Condon
factors.
The above-mentioned approach allows one to connect with one another the
potentials $V(r)$ and $V'(r)$, which might be taken as GMP's, by
identifying them as
members of the set of satellite potentials. The algebraic relation between the
corresponding eigenfunctions $\Psi(r)$ and $\Psi'(r)$, which were
established here,
would then significantly simplify the calculation of Frank-Condon factors.
In such
respect, the $su(1,1)$ subalgebra of~$so(2,2)$ associated with~$g$ looks more
promising than that associated with~$m$, since the operators~$G^{\pm}$, whose
action is illustrated in fig.~2, could describe transitions where the
vibrational
molecular states (characterized by~$n$), and the electronic states (belonging to
definite satellite potentials) change simultaneously. We plan to analyse
this point
further in a forthcoming publication.\par
%
%
\section*{Acknowledgments}
Two of us (ADSM, YuFS) would like to thank Professor E Ley-Koo for his valuable
suggestions and discussions on the subject. This work was supported in part by
CONACYT, Mexico, by the Minist\`ere de l'Education Nationale et de la Culture,
Communaut\'e Fran\c caise de Belgique, and by the Russian Foundation of the
Fundamental Research, Grant No 96-01-01421.\par
%
%
\newpage
\section*{Appendix. SUSYQM analysis and normalization of wave functions}
\renewcommand{\theequation}{A.\arabic{equation}}
\setcounter{section}{0}
\setcounter{equation}{0}
The purpose of the present appendix is to briefly review the SUSYQM approach to
the GMP problem, and to use it to prove eq.~(\ref{eq:normalization}) for
the wave
function normalization coefficient.\par
%
%
Let us consider the Schr\"odinger equation~(\ref{eq:schrodinger-GMP}) for
the GMP
in dimensionless variable~$x$, and denote the corresponding Hamiltonian,
potential, energies, and wave functions by $h^{0)}$, $v^{(0)}$,
$\epsilon^{(0)}_n$,
and $\psi^{(0)}_n(x)$, respectively. In
SUSYQM~\cite{gendenshtein,dabrowska}, the
hamiltonian~$h^{(0)}$ can be written in a factorized form, $h^{(0)} = A^+ A^- +
\epsilon^{(0)}_0$, in terms of the operators $A^{\pm} = \mp d/dx + W(x)$, where
the superpotential~$W(x)$ is related to the ground state wave
function~$\psi^{(0)}_0(x)$ by $W(x) = - d\left(\ln \psi^{(0)}_0(x)\right)
\big/ dx$.
By taking eq.~(\ref{eq:eigenfunction}) for $n=0$ into account, we get
\begin{equation}
  A^{\pm} = \pm y (1+y) \frac{d}{dy} + \alpha_0 + (\alpha_0 - \beta_0) y
  = \mp \frac{d}{dx} + \frac{\alpha_0 e^x - \beta_0}{e^x - 1}.  \label{eq:A}
\end{equation}
\par
%
%
The supersymmetric partner of~$h^{(0)}$ is the Hamiltonian $h^{(1)} = A^- A^+ +
\epsilon^{(0)}_0$. From $h^{(0)}$, $h^{(1)}$, and~$A^{\pm}$, we can form the
generators of an $su(1/1)$ superalgebra, namely the supersymmetric
Hamiltonian~$\cal H$, and supercharges~$Q^{\pm}$, defined by
\begin{equation}
  {\cal H} = \left(\begin{array}{cc}
                          h^{(0)} - \epsilon^{(0)}_0 & 0 \\
                          0 & h^{(1)} - \epsilon^{(0)}_0
                          \end{array}\right) \qquad
  Q^+ = \left(\begin{array}{cc}
                    0 & A^+ \\
                    0 & 0
                   \end{array}\right) \qquad
  Q^- = \left(\begin{array}{cc}
                    0 & 0 \\
                    A^- & 0
                   \end{array}\right)
\end{equation}
respectively, and satisfying the relations
\begin{equation}
  \left[{\cal H}, Q^{\pm}\right] = 0 \qquad \left\{Q^+, Q^-\right\} = {\cal
H} \qquad
  \left\{Q^{\pm}, Q^{\pm}\right\} = 0.
\end{equation}
\par
%
%
From eq.~(\ref{eq:A}), we obtain
\begin{equation}
  h^{(1)} = - \frac{d^2}{dx^2} + v^{(1)}(x)
\end{equation}
where
\begin{equation}
  v^{(1)}(x) = v^{(0)}(x) + 2 \frac{dW(x)}{dx} = k \left(1 -
\frac{b}{e^x-1}\right)^2
  + \frac{2l e^x}{\left(e^x-1\right)^2}
\end{equation}
and $l$ is defined in eq.~(\ref{eq:d-e-l}). The potential $v^{(1)}(x)$ can
be rewritten
as
\begin{equation}
  v^{(1)}(x) = k' \left(1 - \frac{b'}{e^x-1}\right)^2 + R(k',b')
\end{equation}
where
\begin{equation}
  k' = \frac{(kb-l)^2}{kb^2+2l} \qquad b' = \frac{kb^2+2l}{kb-l} \qquad R =
k(k',b') - k'
  = \frac{(k'b'-1+l')^2}{k'b'^2+2-2l'} - k'
\end{equation}
and
\begin{equation}
  l' \equiv \case{1}{2} \left(1 + \sqrt{1+4k'b'^2}\right) = l + 1
\label{eq:lprime}
\end{equation}
hence showing that the GMP is a shape-invariant potential.\par
%
%
The eigenvalues~$\epsilon^{(1)}_n$ of the supersymmetric partner are given by
\begin{equation}
  \epsilon^{(1)}_n = \epsilon^{(0)}_{n+1} = k - \alpha_{n+1}^2
  \label{eq:epsilon1}
\end{equation}
where in the last step, we used eq.~(\ref{eq:epsilon}). Due to the shape
invariance,
they can also be written as
\begin{equation}
  \epsilon^{(1)}_n = k' - \alpha_n^{\prime 2} + R(k',b') = k -
\alpha_n^{\prime 2}
  \label{eq:epsilon1bis}
\end{equation}
where $\alpha'_n$ is defined in terms of $k'$, $b'$, $l'$, and~$n$ in the
same way as
$\alpha_n$ in terms of $k$, $b$, $l$, and~$n$ (see eq.~(\ref{eq:alpha-beta})).
Comparing eq.~(\ref{eq:epsilon1}) with eq.~(\ref{eq:epsilon1bis}), and using
eqs.~(\ref{eq:eq-alpha-beta-1}), and~(\ref{eq:lprime}), lead to the
conclusion that
\begin{equation}
  \alpha'_n = \alpha_{n+1} \qquad \beta'_n = \beta_{n+1}.
\label{eq:alphaprime}
\end{equation}
Hence, the eigenfunctions of the supersymmetric partner are given by
\begin{equation}
  \psi^{(1)}_{n} = \frac{1}{\sqrt{a}} N'_n y^{\alpha_{n+1}} (1+y)^{-\beta_{n+1}}
  {}_2F_1(-n, -n-1-2l, 2\alpha_{n+1}+1,-y)   \label{eq:eigenfunction1}
\end{equation}
where $N'_n$ can be obtained from~$N_n$ by the substitutions $l\to l'$,
$\alpha_n
\to \alpha'_n$, $\beta_n \to \beta'_n$.\par
%
%
From SUSYQM, however, we also know that $\psi^{(1)}_{n-1}$ can be obtained from
the eigenfunction $\psi^{(0)}_n$ of~$h^{(0)}$, corresponding to the same
eigenvalue
$\epsilon^{(0)}_n = \epsilon^{(1)}_{n-1}$, by applying the operator~$A^-$:
\begin{equation}
  \psi^{(1)}_{n-1} = \left(\epsilon^{(0)}_n - \epsilon^{(0)}_0\right)^{-1/2} A^-
  \psi^{(0)}_n = \left(\alpha_0^2 - \alpha_n^2\right)^{-1/2} A^- \psi^{(0)}_n.
  \label{eq:eigenfunction1bis}
\end{equation}
With the help of eqs.~(\ref{eq:A}), (\ref{eq:eigenfunction}), and the relations
\begin{eqnarray}
  \alpha_0 - \alpha_n & = & \frac{n (2\alpha_n+n+2l)}{2l}
\label{eq:alpha-diff} \\
  \alpha_0 - \beta_0 - \alpha_n + \beta_n & = & n
\end{eqnarray}
deriving from~(\ref{eq:eq-alpha-beta-1}) and~(\ref{eq:alpha-beta}), we obtain
\begin{eqnarray}
  A^- \psi^{(0)}_n & = & \frac{1}{\sqrt{a}} N_n y^{\alpha_n}
(1+y)^{-\beta_n} \left[
       - y (1+y) \frac{d}{dy} + \frac{n}{2l} (2\alpha_n+n+2l) + n y\right]
\nonumber \\
  & & \mbox{} \times {}_2F_1(-n, -n+1-2l; 2\alpha_n+1;-y).  \label{eq:A-action}
\end{eqnarray}
By successively using the second relation in eq.~(4.9.7) of
ref.~\cite{infeld}, and
eq.~(1.4.2) of ref.~\cite{slater}, eq.~(\ref{eq:A-action}) can be
transformed into
\begin{eqnarray}
   \lefteqn{A^- \psi^{(0)}_n} \nonumber \\
   & = & \frac{1}{\sqrt{a}} \frac{2\alpha_n+n+2l}{2l} N_n y^{\alpha_n}
       (1+y)^{-\beta_n} \bigl[(n+2l) {}_2F_1(-n, -n+1-2l; 2\alpha_n+1;-y)
\nonumber
       \\
   & & \mbox{} - 2l {}_2F_1(-n, -n-2l; 2\alpha_n+1;-y)\bigr] \nonumber \\
   & = & \frac{1}{\sqrt{a}} \frac{n(2\alpha_n+n+2l)}{2l} N_n y^{\alpha_n}
       (1+y)^{-\beta_n} {}_2F_1(-n+1, -n-2l; 2\alpha_n+1;-y).
       \label{eq:A-actionbis}
\end{eqnarray}
Finally, by combining eqs.~(\ref{eq:eigenfunction1bis}),
(\ref{eq:alpha-diff}),~(\ref{eq:A-actionbis}) with the relation
\begin{equation}
  \alpha_0 + \alpha_n = 2\alpha_n + (\alpha_0 - \alpha_n) = \frac{(n+2l)
  (2\alpha_n+n)}{2l}
\end{equation}
we get
\begin{eqnarray}
  \psi^{(1)}_{n-1} & = &
\left[\frac{n(2\alpha_n+n+2l)}{a(n+2l)(2\alpha_n+n)}\right]
         ^{1/2} N_n y^{\alpha_n} (1+y)^{-\beta_n} \nonumber \\
  & & \mbox{} \times {}_2F_1(-n+1, -n-2l;2\alpha_n+1;-y).
         \label{eq:eigenfunction1ter}
\end{eqnarray}
\par
%
%
We have therefore derived two equivalent expressions~(\ref{eq:eigenfunction1}),
and~(\ref{eq:eigenfunction1ter}) for the wave functions of the GMP
supersymmetric partner. By equating them, we obtain the following recursion
relation for the normalization coefficients
\begin{equation}
  N_n = \left[\frac{(n+2l)(2\alpha_n+n)}{n(2\alpha_n+n+2l)}\right]^{1/2}
N'_{n-1}.
  \label{eq:recursion}
\end{equation}
The recursion starting value is the ground state normalization coefficient,
which
can be expressed in terms of beta or gamma functions, as follows:
\begin{eqnarray}
  N_0 & = & \left[a^{-1} \int_0^{\infty} dy\, y^{2\alpha_0-1}
(1+y)^{-2\beta_0-1}
         \right]^{-1/2} = \left[a^{-1} B(2\alpha_0, 2\beta_0-2\alpha_0+1)\right]
         ^{-1/2} \nonumber \\
  & = & \left[\frac{a \Gamma(2\beta_0+1)}{\Gamma(2\alpha_0)
         \Gamma(2\beta_0-2\alpha_0+1)}\right]^{1/2} = \left[\frac{a
         \Gamma(2\alpha_0+2l+1)}{\Gamma(2\alpha_0) \Gamma(2l+1)}\right]^{1/2}.
         \label{eq:starting}
\end{eqnarray}
It is then a simple matter to prove by using eqs.~(\ref{eq:lprime}),
and~(\ref{eq:alphaprime}), that eq.~(\ref{eq:normalization}) provides the
solution
of eqs.~(\ref{eq:recursion}), and~(\ref{eq:starting}).
\par
%
%

%
%
\newpage
\section*{Figure captions}
\noindent {\bf Figure~1.} Comparison between the GMP (full line), and the Morse
and harmonic oscillator potentials (broken and chain lines respectively). The
parameters of the first two are $a=1$, $r_e$ or $x_e=2.5$, $D=10$, while the
frequency of the latter is $\hbar \omega = \epsilon(1)$.

\noindent {\bf Figure~2.} Effects of the operators $G^{\pm}$ on the
eigenfunctions
of the GMP~($a=1$). In each case, only the first three energy levels are shown.

\end{document}